\documentclass{ifacconf}

\usepackage{graphicx}      
\usepackage{natbib}        
\usepackage{nicefrac}
\usepackage{amsmath,amssymb,amsfonts}
\usepackage{graphicx}   
\usepackage{tikz}
\usepackage{soul}
\usepackage{comment}
\usepackage{mdframed}
\usepackage{framed}

\newcommand{\Rr}{\mathbb{R}}

\newcommand{\mnabla}{\nabla\!}

\newcommand*{\QEDA}{\hfill\ensuremath{\qed}}

\newtheorem{dfn}{Definition}

\begin{document}
\begin{frontmatter}
	
\title{Distributed formation control of networked mechanical systems 
}

\author[First]{N. Javanmardi} 
\author[Second]{P. Borja} 
\author[First]{M. J. Yazdanpanah}
\author[Third]{J. M. A. Scherpen}

\address[First]{Control and Intelligent Processing Center of Excellence, School of Electrical and Computer Engineering, University of Tehran, Tehran, Iran 
	(e-mail:  {najmeh.javanmardi, yazdan}@ut.ac.ir)}
\address[Second]{ Department of Cognitive Robotics, Delft University of Technology, Delft, The Netherlands (e-mail:  l.p.borjarosales@tudelft.nl)}
\address[Third]{Jan C. Wilems Center for Systems and Control, ENTEG, Faculty of Science and Engineering, University of Groningen, Groningen, The Netherlands (e-mail:j.m.a.scherpen@rug.nl)}

\begin{abstract}                
This paper investigates a distributed formation tracking control law for large-scale networks of mechanical systems. In particular, the formation network is represented by a directed communication graph with leaders and followers, where each agent is described as a port-Hamiltonian system with a constant mass matrix. 
Moreover, we adopt a distributed parameter approach to prove the scalable asymptotic stability of the network formation, i.e., the scalability with respect to the network size and the specific formation preservation. A simulation case illustrates the effectiveness of the 
proposed control approach.
\end{abstract}

\end{frontmatter}
\section{Introduction}
Recently, distributed formation controllers for networked systems have received considerable attention due to numerous applications such as platoons of autonomous vehicles in transportation systems  \cite{dunbar2011distributed},  microgrids \cite{yazdanian2014distributed}, and multi-robot formation maneuvers \cite{ren2008distributed}.
These control strategies have proven successful in reducing heavy computational burden,  communication range, and unreliability in large-scale networked systems by using only local information of the neighboring agents.  However, unstable behaviors and error amplification in large-scale networked systems may occur as the number of agents increases \cite{hao2011stability,barooah2009mistuning}. Hence, the stability and performance analysis of large-scale networked systems have been relevant topics in the recent literature, e.g., \cite{hao2011stability,dashkovskiy2020stability,besselink2018scalable}. In particular, there exists a vast literature on large-scale vehicular formation, e.g., \cite{dunbar2011distributed,hao2011stability,barooah2009mistuning,ghasemi2013stable,knorn2015scalability}.

In \cite{barooah2009mistuning, meurer2012control}, the authors adopt approaches that rely on approximating the dynamics of the agents via partial differential equations (PDEs) to analyze the formation problem for large-scale interconnected systems. An advantage of these methods over ordinary differential equation (ODE) models is that in PDE-based models, the number of agents is a parameter that does not affect the dimension of the networked system. Consequently, increasing the size of the network does not increase the dimension of the PDE model. Moreover, PDE models provide a 2-dimensional framework for better insight into the network performance in both space and time.

The port-Hamiltonian modeling framework is suitable to represent a broad range of physical systems while underscoring the role of the energy, dissipation, and interconnection pattern in the behavior of these systems (see \cite{duindam}). As exposed in \cite{ortega2002stabilization}, a control design methodology adequate to stabilize complex mechanical systems represented as pH systems is the so-called interconnection and damping assignment passivity-based control (IDA-PBC) approach. In \cite{valk2018distributed,tsolakis2021distributed}, the authors adopt this approach to solve the stabilization problem for a class of networks of mechanical systems. Regarding the trajectory-tracking problem, the notion of contractive pH systems is used in \cite{yaghmaei2017trajectory} to develop a tracking version of IDA-PBC, named timed IDA-PBC (tIDA-PBC). Recently, in \cite{javanmardi2020spacecraft}, this method has been used to address the spacecraft formation flying (SFF) problem.

This paper proposes a distributed nonlinear control scheme that ensures tracking of the reference trajectories while maintaining the desired formation geometry for a class of large-scale networks of mechanical systems. To this end, the agents are modeled as pH systems and the tIDA-PBC notion is used to develop a distributed tracking formation law. Furthermore, a leader-follower architecture is adopted, where the leaders have access to the trajectory information, and the followers use only local information to achieve the desired formation. Then, a PDE-based method is used to investigate the scalable asymptotic stability (SAS) of the networked system. Notably, this property guarantees that the performance of the networked system is independent of the size of the network. Thus, the main contributions of this paper are:
\begin{itemize}
	\item  We solve the formation stabilization and trajectory-tracking problems for pH systems simultaneously without changing the controller structure for a class of large-scale networks of mechanical systems. To this end, we propose a distributed control law that requires only local information of the neighbors. 
	\item We derive a PDE approximate model of the closed-loop system that gives a 2-dimensional perspective of the network.
	\item  The notion of SAS is introduced to investigate the scalability of the networked systems with respect to the network size.
\end{itemize}

We stress that compared with the centralized controller reported in \cite{javanmardi2020spacecraft}, this work proposes a distributed controller suitable for solving the SFF problem. Moreover, in contrast to \cite{valk2018distributed}, \cite{tsolakis2021distributed}, where only stabilization is investigated, this paper addresses the trajectory-tracking problem while assessing the performance of the large-scale networked system. In this regard, this work differs from \cite{dunbar2011distributed, barooah2009mistuning,hao2011stability,ghasemi2013stable,knorn2015scalability} by considering non-constant velocity references and the mechanical model of the networked system. Additionally, contrary to the mentioned references, we use the SAS concept to analyze the PDE-based approximation of the closed-loop network.

\textbf{Notation:} Given the matrix $A \in \mathbb{R}^{n \times n}$, its symmetric part is denoted by $\mbox{sym}(A)\triangleq\frac{1}{2}(A+A^\top)$. Consider $x\in\mathbb{R}^{n}$ and a continuously differentiable function of $H(x):\mathbb{R}^n \longrightarrow \mathbb{R}$. Then, $\nabla H(x)$ is defined as $[\,\frac{\partial H}{\partial x_1}, \frac{\partial H}{\partial x_2}, \dots \frac{\partial H}{\partial x_n}]\,^\top$ and ${\nabla}^2 H$ denotes the \textit{Hessian} of $H(x)$. The symbols $I_{n_i}$ and $0_{n_i}$ are used for the identity matrix of dimension ${n_i}$ and the zero matrix of dimension ${n_i}\times {n_i}$, respectively. The Euclidean norm of $x$ is indicated as $\Vert \cdot \Vert$. The absolute value of a real number $c \in \mathbb{R}$ is denoted by $\vert\cdot\vert$.
\vspace{-4pt}
\section{Preliminaries}
\vspace{-2pt}
\subsection{Timed IDA-PBC}
\vspace{-1pt}
\label{sec:p1}
Consider the following input-state-output pH system 
\begin{equation}
\label{eq:PHOL}
\begin{split}
&\dot{x}=(J(x)-R(x))\nabla H(x) + g(x) u,\\&
y=g^\top(x)\nabla H(x),
\end{split}
\end{equation}
where  $x(t)\subseteq\mathbb{R}$ is the state of the system; the interconnection matrix $J:\Rr^n \to \Rr^{n \times n}$ is \textit{skew-symmetric}; the damping matrix $R:\Rr^n \to \Rr^{n \times n}$ is \textit{positive semi-definite}; $H: \Rr^n \to \Rr$ denotes the Hamiltonian function; $g:\Rr^n \to \Rr^{n \times m} $ is the input matrix; and $u,y\in\Rr^m$ are the control input and the passive output, respectively.

To introduce the tIDA-PBC method, we need the following definition:

\begin{dfn}
	\label{definition1}
	Consider an open subset of $\Rr_{+}$ denoted by $\mathbb{I}$. Then, the trajectory $x^\star(t)$ is a feasible trajectory of \eqref{eq:PHOL} if there exists $u^\star(t)$ such that 
	\begin{equation*}
	\dot{x}^\star=(J(x^\star(t))-R(x^\star(t))\nabla H(x^\star(t))+g(x^\star(t))u^\star(t) 	
	\end{equation*}
	for $t\in \mathbb{I}$.
	$\QEDA$
\end{dfn}
The following theorems present the notion of contractive pH systems and the tIDA-PBC approach. We refer readers to  \cite{yaghmaei2017trajectory} for a thorough exposition on tIDA-PBC and the proof of Theorems \ref{Th1} and \ref{Th2}. 

\begin{thm}
	\label{Th1}
	Consider the following system:
	\begin{equation}
	\label{closed sys}
	\begin{split}
	\dot{x}(t)=F_d \mnabla H_d(x, t),
	\end{split}
	\end{equation}
	with $F_d\triangleq J_d-R_d$, where $J_d=-J^\top_d$ and $R_d=R^\top_d\succeq 0$ are the desired interconnection and damping matrices, respectively, and the desired Hamiltonian function  $H_d(x,t): \Rr^n \times \Rr_+  \to \Rr$, which satisfies the following condition:
	\begin{equation}
	\label{cond3}
	\begin{split}
	&\alpha I\prec \nabla^2 H_d(x, t) \prec \beta I,\quad \forall x \in D_T\subset \Rr^n 
	\end{split}
	\end{equation}
	for constants $0<\alpha<\beta$. Moreover, suppose that all eigenvalues of  $F_d$ have strictly negative real parts. Then, the system \eqref{closed sys} is contractive on an open subset of $D_T$ if the following matrix has no eigenvalues on the  imaginary axis for a positive constant $\varepsilon$:
	\begin{equation}
	\label{eq:N}
	N = \begin{bmatrix}
	F_d & \eta F_dF_d^\top\\ -(\eta+\varepsilon)I & -F_d^\top
	\end{bmatrix},
	\end{equation}
	where $\eta\triangleq 1-\frac{\alpha}{\beta}$.
\end{thm}
\vspace{-6pt}	
The contractivity property guarantees that all system trajectories converge exponentially to each other. Thus, if  $x^\star(t)$ is a trajectory of a contractive system, all other trajectories converge exponentially to $x^\star(t)$. The following theorem represents the tIDA-PBC method. 

\begin{thm}[timed IDA-PBC]
	\label{Th2}
	Consider the desired trajectory $x^\star(t)$ as a feasible trajectory of system \eqref{eq:PHOL}. Assume that there exist $F_d$ and $H_d$ satisfying the contractivity conditions of Theorem \ref{Th1} and the following matching equation: 
	\begin{equation}
	\label{eq:mathcingT}
	\begin{split}
	g^\perp \{ (J(x)-R(x))\nabla H(x)-F_d\mnabla H_d(x,t)\}=0,
	\end{split}
	\end{equation}
	where $g^\perp(x)$ is the full rank left annihilator of $g(x)$. If $F_d$ and $H_d$ satisfy
	\begin{equation}
	\label{eq:condT}
	\dot{x}^\star(t)=F_d \mnabla H_d(x^\star(t), t),
	\end{equation}
	then, the controller \begin{equation}
	\label{eq:UT}
	\begin{split}
	u = g^{\dagger}(x)\{ F_d\mnabla H_d(x,t)
	-(J-R)\mnabla H(x)\}. &
	\end{split}
	\end{equation}
	where $g^{\dagger}\triangleq(g^{\top}g)^{-1}g^{\top}$,  makes the system \eqref{eq:PHOL} a local exponential tracker for $x^\star(t)$. 
\end{thm}
\vspace{-6pt}
\section{Problem Statement and Network control}
\vspace{-6pt}
This section focuses on the modeling,  control objectives, and control design for the networked system.
\vspace{-6pt}
\subsection{Network modeling and control objectives}
\label{model-obj}
\vspace{-6pt}
Consider a large-scale networked system composed of $N \in \mathbb{N}$ agents (subsystems), each agent $i$ represented by a (possibly damped) fully actuated mechanical system with $n_i$ degrees of freedom and \textit{positive definite} constant mass matrix $M_0\in\Rr^{n_i \times n_i}$. Hence, the pH representation of each agent is
\begin{equation} 
\label{open-loop-2}
\arraycolsep=0.8pt
\def\arraystretch{1.2}
\begin{array}{rcl}
\begin{bmatrix}
{{{\dot q}_i}} \\ 
{{{\dot p}_i}} 
\end{bmatrix} &=& \begin{bmatrix}
0_{n_i}&{{I_{n_i}}} \\ 
{ - {I_{n_i}}}&-\mathcal{R}_i
\end{bmatrix}\begin{bmatrix}
{{\nabla _{{q_i}}}{{H}_i(q_i,p_i)}} \\ 
{{\nabla _{{p_i}}}{{H}_i(q_i,p_i)}} 
\end{bmatrix}+\begin{bmatrix}
0 \\ 
I_{n_i}
\end{bmatrix}u_i,\\[0.4cm]
H_i(q_i,p_i) &=& \frac{1}{2}{p_i^\top}M^{-1}_0 p_i+\mathcal{U}_i(q_i).
\end{array}
\end{equation}
where $q_i, p_i, u_i\in {\mathbb{R}^{n_i}}$ are the position, the momenta, and the input of each agent $i$, respectively. Moreover, $\mathcal R_i\in\Rr^{n_i \times n_i}$ is the \textit{positive semi-definite} damping matrix and $\mathcal{U}_i:\Rr^{n_i}\to\Rr$ denotes the potential energy of the agent $i$. We stress that the agents may have non-identical potential energies. Thus, the network may be composed of heterogeneous agents.

We assume that it is possible to exchange information among the subsystems in the network. In particular, the information flow among agents in the network is modeled by a communication graph $G(\mathcal{N},\mathcal{E})$ with a set $\mathcal{N}$ of nodes, which represent the agents, and a set $\mathcal E \subseteq \mathcal N \times \mathcal N$ of edges. Two nodes $i$ and $j$ are called neighbors if $(i,j)\in \mathcal{E}$. Neighboring between the nodes $i$ and $j$ is  indicated by $i \sim j$. In this paper, we consider connected and directed graphs, where a path (i.e., a sequence of edges) exists between every pair of nodes, and the information flow among agents is assumed unidirectional. Moreover, we assume that the communication graph has no self-loops, i.e., $(i,i) \notin \mathcal E$. The agents of the network are leaders or followers , where $\mathcal{B}_L$ and $\mathcal{B}_F$ denote the non-empty sets of leaders and followers, respectively, and $\mathcal{B}_L\cup\mathcal{B}_F=\mathcal{N}$. Furthermore, the constant vector $\Delta_{i,j}\in {\mathbb{R}^{n_i}}$, defined as $\Delta_{i,j}\triangleq q_i(t)- q_j(t)$, indicates the desired formation geometry between the pair of agents $(i,j)$. 
Then, the control objective is to design a distributed control that ensures tracking of the reference trajectories while each follower agent maintains the desired formation geometry and guarantees the stability of the network independently of its size. To this end, we assume that each leader knows its reference trajectory position $q_{r,i}(t)$ and momenta $p_{r,i}(t)$, for $i\in \mathcal{B}_L$, and followers have information of their neighbors and know the inter-agents difference $\Delta_{i,j}$. Hence, the control objectives take the form
\begin{equation}
\label{OBJ}
\begin{array}{rcll}
\mathop {\lim }\limits_{t \to \infty } \left\| {{q_i(t)} - {q_j(t)} - {\Delta_{i,j}}} \right\|& = 0,&\forall{i,j}|\, i\sim j,\,\,\,i\in \mathcal{B}_F\\
\mathop {\lim }\limits_{t \to \infty } \Vert {{q_i(t)} - q_{r,i}(t)} \Vert& = 0,&\forall i\, \in \mathcal{B}_L\\
\mathop {\lim }\limits_{t \to \infty } \Vert{{\dot{q}_i(t)} - \dot{q}_{r,i}(t)} \Vert& = 0,&\forall i\, \in \mathcal{B}_L, \mathcal{B}_F.
\end{array}
\end{equation}
In addition, we require the network to exhibit the scalable asymptotic stability (SAS) property. The concept of SAS is instrumental in studying the effect that variations in the size of the network (i.e., number of agents) have on the performance of the networked system.

\begin{dfn}
	\label{def:SALS1}
	A networked system with $N \in \mathbb{N}$ agents is said to be SAS if, for any $\varepsilon > 0$, there exists $\delta  = \delta (\varepsilon ) > 0$ such that 
	\begin{equation}
	\label{Cond:SLS1}
	\max\limits_{i \in  \mathcal N} \Vert x_i(t_0)\Vert < \delta \implies  \max\limits_{i \in  \mathcal N} \Vert x_i(t)\Vert < \varepsilon
	\end{equation} 
	where $x_i$ is the state of each agent $i$, for all $N \in \mathbb{N}, t\ge t_0$,
	and $\delta$ can be chosen such that 
	$\lim\limits_{t \to \infty } \max\limits_{i \in  \mathcal N} \Vert x_i(t)\Vert= 0.
	$ $\QEDA$
\end{dfn}
According to the notion of SAS, the state trajectories of the networked system remain bounded for any $N\in \mathbb{N}$, i.e., boundedness is independent of the size of the network. Hence, the performance of the networked system is scalable. Consequently, the stability of the networked system is not jeopardized by adding or removing agents.
\vspace{-7pt}
\subsection{Local distributed control}
\label{dis-control}
\vspace{-7pt}
In this section, we propose a local distributed controller for the networked system. To this end, we adopt the tIDA-PBC approach.
\vspace{-7pt}
\subsubsection{Tracking control law for each agent:}
Consider the agent $i$ with dynamics \eqref{open-loop-2}. To achieve contractivity, the closed-loop system is designed to satisfy the conditions established in Theorem \ref{Th1}. Hence, the target dynamics are 
\begin{equation} 
\label{closed-loop-2}
\begin{bmatrix}
{{{\dot q}_i}} \\ 
{{{\dot p}_i}} 
\end{bmatrix}= \begin{bmatrix}
0_{n_i}&{{I_{n_i}}} \\ 
- {{I_{n_i}}}&  \bar{\mathcal{J}}_i - \bar{\mathcal{R}}_i 
\end{bmatrix} \begin{bmatrix}
{{\nabla _{{q_i}}}{{H_{d,i}(q_i,p_i,t)}}} \\ 
{{\nabla _{{p_i}}}{{H_{d,i}(q_i,p_i,t)}}}
\end{bmatrix},
\end{equation}
where $\bar{\mathcal{J}}_{i}\in\Rr^{n_i\times n_i}$ and $\bar{\mathcal{R}}_i\in\Rr^{n_i\times n_i}$ are \textit{skew-symmetric} and \textit{positive definite}, respectively.
Then, these matrices must be chosen such that 
\begin{equation}
\label{N}
F_d := \begin{bmatrix}
0_{n_i}&{{I_{n_i}}} \\ 
- {{I_{n_i}}}&  \bar{\mathcal{J}}_i - \bar{\mathcal{R}}_i 
\end{bmatrix}
\end{equation}
is Hurwitz and the condition on \eqref{eq:N} is satisfied.
Moreover, the desired Hamiltonian function is described as
\begin{equation}
\label{Hd-2}
H_{d,i}(q_i,p_i,t)=\frac{1}{2}p_i^{\top}M^{-1}_{0}p_i+\frac{1}{2}(q_i-L_i(t))^{\top}K_i(q_i-L_i(t)), 
\end{equation}
where $K_i\in\Rr^{n_i\times n_i}$ is \textit{positive definite}. Thus, \eqref{cond3} holds. Then, to obtain a system of the form \eqref{eq:condT}, $L_i(t)$ is chosen as 
\begin{equation}
\label{Lgain-2}
L_i(t)=q_i^\star(t)-K_i^{-1}\big((\bar{\mathcal{J}}_{i}-{\bar{\mathcal{R}}}_i)M_{0}^{-1}{p}_i^\star(t)-\dot{p}^\star_i(t)\big).
\end{equation}
Therefore, the target system \eqref{closed-loop-2} is contractive. Note that, since the agents are fully actuated, the matching conditions are trivially satisfied. Consequently, the control law obtained from the tIDA-PBC method is
\begin{equation}
\label{controller-2}
\begin{split}
u_i&=
\nabla_{q_i} \mathcal{U}_i(q_i) - K_i(q_i-L_i(t)) \\
&\quad  + {\mathcal{R}_i} M^{-1}_i {p}_i +(\bar{\mathcal{J}}_{i}-{\bar{\mathcal{R}}}_i)M_{i}^{-1}{p}_i.
\end{split}
\end{equation}
\vspace{-7pt}
\subsubsection{Admissible controller:} 
\vspace{-7pt}
Each agent in the network must track an appropriate, feasible trajectory to achieve the control objectives established in Section \ref{model-obj}. To this end, the control structure \eqref{controller-2} is implemented for all agents. In particular, for leaders, the feasible desired trajectories are 
\begin{equation}
\label{Lrefrence}
x^\star_i(t)=\begin{bmatrix} {{{q}^\top_{r,i}}(t)},p^\top_{r,i}(t)\end{bmatrix}^\top,\quad\forall i\, \in \mathcal{B}_L.
\end{equation}
where $p_{r,i}(t)\triangleq M_0{\dot{q}}_{r,i}(t)$. Concerning the followers, the desired trajectories are proposed as an average of the neighbors information, i.e.,
\begin{equation}
\label{Frefrence}
x^\star_i(t)=\frac{1}{Q_i} \sum\limits_{j \in \mathcal{N},\,{i} \sim {j}}\begin{bmatrix} {{q}^\top_{j}(t)+\Delta_{i,j}^\top},{{p_i^{\star}}^\top(t)}\end{bmatrix}^\top,\quad\forall i\, \in \mathcal{B}_F
\end{equation}
where 
$p_i^{\star}(t)\triangleq M_0{{\dot{q}^\star_{i}}(t)}$ and 
$Q_i$ is the number of neighbors of agent $i$. We remark that the proposed distributed control protocol for a follower depends on the local positions, velocities, and accelerations of its neighbors. From a practical perspective, the accelerations can be calculated via numerical differentiation of the velocities. Note that only the leaders have access to the reference information. Here, we consider an ideal communication network, i.e., no communication delay or failure is considered.
Substitute \eqref{Lgain-2} in \eqref{closed-loop-2} to obtain
\begin{equation}
\label{closed-loop1}
\begin{split}
\dot p_i - \dot p_i^\star  &= - K_i (q_i-q^\star_i) + (\bar{\mathcal{J}}_{i}-{\bar{\mathcal{R}}}_i)M^{-1}_{0} (p_i-p_i^\star),
\end{split}
\end{equation}
where $\dot{q}_i=M_0^{-1}p_i$.
Let $q_\ell(t)$ and $p_\ell(t)$ denote the position and momentum of the leader, respectively. Then, the desired trajectory of agent $i$ is defined by the trajectory of the corresponding leader agent and the desired formation geometry, i.e, $q^{ref}_i(t)=q_\ell(t)+\Delta_{i,\ell}$. Hence, the desired momentum of the formation can be defined as $p^{ref}_i=p_\ell(t)$. Thus, the position and momentum errors of agent $i$ are
\begin{equation}
\label{Error}
\tilde{q}_i\triangleq q_i-q_i^{ref},\quad \tilde p_i\triangleq M_0\dot{\tilde q}_i.
\end{equation}
Hence, considering $\Delta_{i,k}=\Delta_{i,j}+\Delta_{j,k}$ for every triplet $(i,j,k)$, and substituting \eqref{Error} and \eqref{Frefrence} into  \eqref{closed-loop1}, the error dynamics of each agent $i$ are given by
\begin{equation}
\label{closed-loop-tilde}
\begin{split}
\dot{\tilde p}_i=-K_i(\tilde q_i-\tilde q^\star_i)+{\dot{\tilde{p}}}^\star_i-(\bar{\mathcal{J}}_{i}-{\bar{\mathcal{R}}}_i)M^{-1}_{0}(\tilde{p}^\star_i-\tilde{p}_i),
\end{split}
\end{equation}
with
\renewcommand{\arraystretch}{.5}
\begin{equation}
\label{reftilde}
\begin{split}
\tilde q^\star_i\triangleq\frac{1}{Q_i} \sum\limits_{j \in \mathcal{N},\,{j} \sim {i}} \tilde q_j,\,\,\,\,\quad 
\tilde p^\star_i\triangleq\frac{1}{Q_i} \sum\limits_{j \in \mathcal{N},\,{j} \sim {i}} \tilde p_j.
\end{split}
\end{equation}
where $\tilde q^\star_i=\tilde p^\star_i = 0$, if the agent $i$ is a leader. Similarly, due to the equality between the desired trajectory and the trajectory of a leader agent, $\tilde q_i=0$ if $i$ is a leader agent.

\begin{rem}
	The stability and scalability performance of the network is affected by the control architecture and communication graph topology. Therefore, the networked system may be unstable even if every agent is stable. 
\end{rem}

\begin{rem}
	The concept of SAS, given in Definition \ref{def:SALS1} presents two perspectives of the networked systems: $i)$ the temporal perspective related to the convergence of states through the time,  and $ii)$ the spatial perspective, which considers the boundedness of states with respect to the increasing number of agents.
	In this context, a two-dimensional description of the networked system can be developed referring to the fact that the system depends on two independent variables, i.e., time and the position/location within the network.
	Therefore, a methodology based on the PDE model,
	which changes the system description into a two-dimensional system, provides a more exact insight into the behavior and characterization of the closed-loop networked system.
\end{rem}
\section{PDE-based model of the networked system}
\label{PDE}
In this section, the PDE model and SAS property of the networked system are discussed. 
\subsection{PDE model of the controlled agents in the network}
\label{PDEmodel}
Combining equations \eqref{closed-loop-tilde} and \eqref{reftilde}, multiplying both sides by $Q_i$, and rewriting the equation in terms of $\tilde q_i$, we obtain
\begin{equation}
\label{closed-dis}
\begin{split}
&{Q_i}\ddot{\tilde{q}}_i=-M_0^{-1}K_i\left({Q_i}{\tilde{q}}_i-\sum\limits_{j \in \mathcal{N},\,{j} \sim {i}}{\tilde{q}_j}\right)\\&+\sum\limits_{j \in \mathcal{N},\,{j} \sim {i}}{{\ddot{\tilde{q}}_j}}+M_0^{-1}(\bar{\mathcal{J}}_{i}-{\bar{\mathcal{R}}}_i)\left({Q_i}\dot{\tilde{q}}_i-\sum\limits_{j \in \mathcal{N},\,{j} \sim {i}}{{\dot{\tilde{q}}_j}}\right).
\end{split}
\end{equation}
This finite-dimensional error dynamics, for $i\in \mathcal{N}$, completely describes the dynamical position behavior of the controlled network of interconnected agents.

The following definition is necessary to present the results of this section.

\begin{dfn}	
	\label{definition3}
	The $\mathcal{D}$-dimensional mesh-like architecture is a graph which is defined in the space $\mathbb{R}^{D}$ with a Cartesian reference frame whose axes are denoted by $z^k (k=1,...,D)$. The number of agents at each direction is indicated by $r^k$ ($N=r^1\times...\times r^D$). By considering the graph as a unit D-cell $[0,1]^D$, the desired location of each agent $i$ in the graph coordinate can be represented at the location $[(i^1-1)\delta^1,...,(i^D-1)\delta^D]^\top$
	, which indicates its spatial coordinate, where $i^k\in \left\{1,...,r^k\right\}$ and the discretization step-size is $\small {\delta^k=\dfrac{1}{r^k-1}}$ for all $k\in \left\{1,...,D\right\}$.$\QEDA$ 
\end{dfn}
Henceforth, we consider mesh-like architectures as communication graphs. 
Therefore, we can recast the summation of positions of neighboring agents in the $\mathcal{D}$-dimensional graph as follows 
\begin{equation}
\begin{split}
\sum\limits_{j \in \mathcal{N},\,{j} \sim {i}}{\tilde{q}_j}=\sum\limits_{j \in \mathcal{N}_i^{1}}{{\tilde{q}_{j}}}+ ... +\sum\limits_{{j} \in \mathcal{N}_i^{D}}{{\tilde{q}_{j}}},
\end{split}
\end{equation}
where $\mathcal{N}_i^{k}$ denotes a set of neighboring nodes of the $i$th agent in the $k$th direction with $Q_i^k$ agents ($Q_i=Q_i^1+...+Q_i^D$). Then, \eqref{closed-dis} can be reformulated as follows
\begin{equation}
\begin{split}
&{Q_i}\frac{\partial^2 \tilde{q}_i(t)}{\partial t^2}=-M_0^{-1}(\bar{\mathcal{J}}_{i}-{\bar{\mathcal{R}}}_i)\left(\sum\limits_{k=1}^{D}\delta^k \frac{{{\dot {\tilde{q}}_j(t)}|_{{j} \in \mathcal{N}_i^{k}}-\dot {\tilde{q}}_i}(t)}{\delta^k}\right)\\&
+M_0^{-1}K_i\left(\sum\limits_{k=1}^{D}{\delta^k} \frac{\tilde{q}_j(t)|_{{j} \in \mathcal{N}_i^{k}}-\tilde{q}_i(t)}{\delta^k}\right)
+\sum\limits_{k=1}^{D}{{\ddot{\tilde{q}}_j(t)}|_{{j} \in \mathcal{N}_i^{k}}}\,.
\end{split}
\end{equation}
\begin{figure}
	\centering
	\includegraphics[width=0.6\columnwidth]{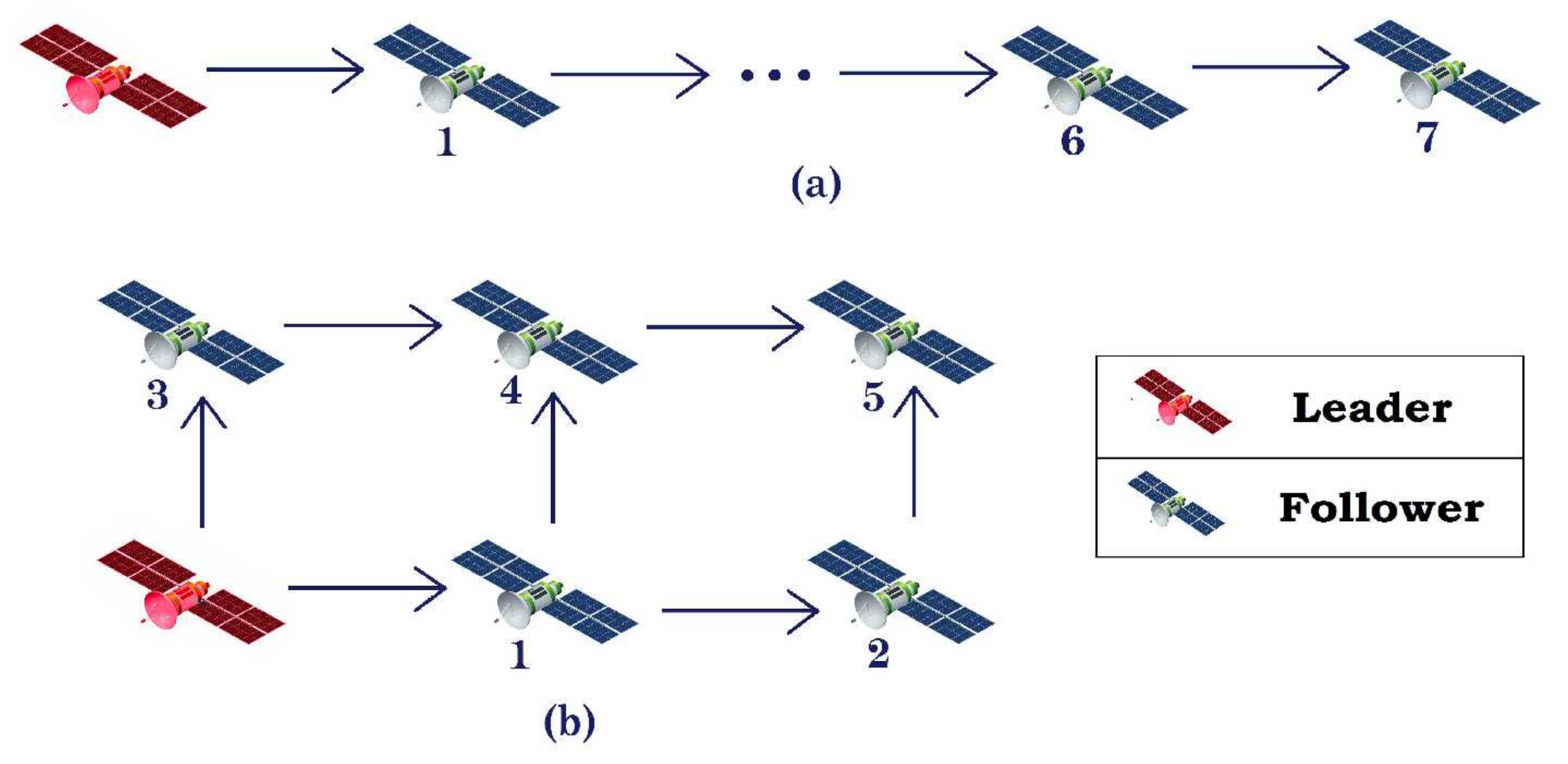}
	\centering
	\caption {Two different communication graphs with (a) 1- dimensional and (b) 2-dimensional.}
	\label{Communication graph}
\end{figure}
\begin{rem}
	\label{remark2}
	We consider a directed and $\mathcal{D}$-dimensional mesh-like graph such that each agent receives information from its immediate predecessor in each direction $z^k$, and one leader in the position $\vec{z}=[0,...,0]^\top$. Examples of possible communication architectures in dimensions $D=1$ and $D=2$ are depicted in Fig. \ref{Communication graph}.
\end{rem}

Consider the positions of each agent as evaluations of the position field over the grid of desired locations as , i.e.,
$\tilde{q}(\vec{z},t):[0,1]^D\times[0,\infty)\rightarrow\Rr^{n_i}$, which satisfies $\tilde{q}_i(t)=\tilde{q}(\vec{z},t)|_{\vec{z}=[(i^1-1)\delta^1,...,(i^D-1)\delta^D]^\top}$. Given the unidirectional information flow mentioned in Remark \ref{remark2}, the following finite difference approximation is considered for $k\in \left\{1,...,D\right\}\ $
\begin{equation}
\label{finitediffrence}
\frac{1}{\delta^k}\,(\tilde{q}_{j}|_{{j} \in \mathcal{N}_i^{k}}-\tilde{q}_{i})=\left[\frac{\partial \tilde{q}(\vec{z},t)}{\partial{z^k}}  \right]_{\vec{z}=[(i^1-1)\delta^1,...,(i^D-1)\delta^D]^\top}.
\end{equation}
Consider the design matrices as distributed field variables
$K,\bar{\mathcal{J}},\bar{\mathcal{R}}:\,\,\,{\left[ {0,1} \right]^{{D}}}\to \mathbb{R}^{{n_i}\times {n_i}}$. Then, we have
\begin{equation}
\label{fuctionPDE}
\begin{split}
&K_i=K(\vec{z}),\,\,\bar{\mathcal{J}}_{i}=\bar{\mathcal{J}}(\vec{z}),\,\ \bar{\mathcal{R}}_i=\bar{\mathcal{R}}(\vec{z}),\,\,
\end{split}
\end{equation}
at $\,\vec{z}=[(i^1-1)\delta^1,...,(i^D-1)\delta^D]^\top$. Hence, we formulate the following PDE
\begin{equation}
\label{PDE-model}
\begin{split}
&F_1(\vec{z})\left(\sum\limits_{k=1}^{D}\delta^k \frac{\partial \tilde{q}(\vec{z},t)}{\partial z^k}\right)+F_2(\vec{z})\left(\sum\limits_{k=1}^{D}\delta^k \frac{\partial}{\partial t} \frac{\partial \tilde{q}(\vec{z},t)}{\partial z^k}\right)\\&+\sum\limits_{k=1}^{D}\delta^k \frac{\partial^2}{\partial t^2} \frac{\partial \tilde{q}(\vec{z},t)}{\partial z^k}=0,
\end{split}
\end{equation}
where
$F_1(\vec{z}):= M_0^{-1}K(\vec{z}),
F_2(\vec{z}):= -M_0^{-1} (\bar{\mathcal{J}}(\vec{z})-{\bar{\mathcal{R}}(\vec{z})}),$
under the Dirichlet-Neumann boundary conditions
\begin{equation}
\label{eq:BC}
\begin{split}
&\tilde q(\vec o,t) = 0,\quad
{\left. {\frac{{\partial \tilde q}}{{\partial {z^k}}}(\vec z,t)} \right|_{{z^k} = 0\,\,\,or\,\,1}} = \bar c_k,\quad k \ge 1 ,\,\vec z \neq 0,
\end{split}
\end{equation}
where $\bar c_k$ is the constant parameter. Thus, the closed-loop dynamics \eqref{closed-dis} can be realized with \eqref{finitediffrence} and \eqref{fuctionPDE} as a finite-difference approximation of \eqref{PDE-model} in the grid points (mesh-like graph in \eqref{definition3}). The approximation improves when the number of agents in each directions, i.e., $r^k$, increases. Furthermore, the initial condition $\tilde q (\cdot,t_0)$ is determined by the difference between the initial position of the agents and their positions on the desired formation, which is
\begin{equation}
\label{eq:IC}
\Vert\tilde q (z,t_0)\Vert<\alpha, \quad \alpha>0 .
\end{equation}
\subsection{Stability of the network with identical controllers }
We consider that the local controllers \eqref{controller-2} have identical design matrices for each agent, i.e., $K(\vec {z})=K_0$, $\bar{\mathcal{J}}(\vec {z})=\bar{\mathcal{J}}_0$, $\bar{\mathcal{R}}(\vec {z})=\bar{\mathcal{R}}_0$. Then, \eqref{PDE-model} reads
\begin{equation}
\label{eq:PDE-homo-model}
\begin{split}
&\sum\limits_{k=1}^{D}\left(\bar F_1\delta^k \frac{\partial }{\partial z^k}+\bar F_2 \delta^k \frac{\partial}{\partial t} \frac{\partial}{\partial z^k}+\delta^k \frac{\partial^2}{\partial t^2} \frac{\partial}{\partial z^k}\right)\tilde{q}(\vec{z},t)=0,
\end{split}
\end{equation}
where $\bar F_1\triangleq M_0^{-1}K_0$, $F_2\triangleq -M_0^{-1} (\bar{\mathcal{J}_0}-\bar{\mathcal{R}}_0)$.
\begin{thm}
	Consider a network of agents with dynamics \eqref{open-loop-2}; the distributed controllers \eqref{Lgain-2}, \eqref{controller-2} and \eqref{Frefrence}; and the information flow architecture given in Definition \ref{definition3}.
	The following statements hold:
	\begin{itemize}
		\item [3.1)]
		The closed-loop networked system \eqref{closed-loop-tilde} is realized as a spatial discretization of the PDE model \eqref{PDE-model} under the boundary conditions \eqref{eq:BC} and the initial condition \eqref{eq:IC}.
		\item [3.2)]
		By considering constant and identical mass and design matrices, the closed-loop networked system \eqref{closed-loop-tilde} is SAS and the objectives \eqref{OBJ} are achieved.
	\end{itemize}
	
	\begin{pf}  To prove the statement 3.1), we rewrite \eqref{closed-loop-tilde} and \eqref{reftilde} as in \eqref{closed-dis}. Then, we obtain the PDE model \eqref{PDE-model} from \eqref{finitediffrence} and \eqref{fuctionPDE}.
		To prove the statement 3.2) and the SAS property of the network, note that every $\tilde{q}_i(t)$ is bounded for any  $N\in\mathbb{N}$ and $t\geq 0$. Therefore, we analyze the stability of the PDE model  \eqref{eq:BC}, \eqref{eq:IC}, and \eqref{eq:PDE-homo-model}. Regarding the invariant separation method, the solution of the linear PDE equation \eqref{eq:PDE-homo-model} with constant coefficients---considering functions $ T(t)\in {\mathbb{R}^{n_i}}$ and 
		$X_k(z^k):\,\,{\left[ {0,1} \right]}\to \mathbb{R}$---can be presented as
		\begin{equation}
		\label{eq:seperation-Var}
		\tilde{q}(\vec{z},t)=X_1(z^1)\cdot X_2(z^2)...X_D(z^D)\cdot T(t)=\prod\limits_{k=1}^{D} {X_k(z^k)}\cdot T(t)\,.
		\end{equation}
		Thus, substituting \eqref{eq:seperation-Var} into \eqref{eq:PDE-homo-model}, we get
		\begin{equation}
		\label{eq:PDE-seperable}
		\begin{split}
		&\left(\sum\limits_{k=1}^{D}{\frac{dX_k(z^k)}{dz^k}\cdot\frac{\delta^k}{X_k(z^k)}}\right)\cdot\left(\bar F_1T(t)+\bar F_2\dot T(t)+\ddot T(t)\right)=0.
		\end{split}
		\end{equation}
		Therefore, it suffices to investigate the stability of system \eqref{eq:PDE-seperable}. Moreover, from \eqref{eq:seperation-Var}, we conclude that
		\begin{equation}
		\label{eq:sep-Var-norm}
		\Vert \tilde{q}(\vec{z},t)\Vert=\prod\limits_{k=1}^{D}  \vert{X_k({z}^k)}\vert\cdot \Vert T(t)\Vert,
		\end{equation}
		which shows that the boundedness of $\tilde q(\vec{z},t)$ is specified by the bounds of $X_k({z}^k)$ and $T(t)$. Then, consider
		\begin{equation}
		\label{eq:second-order}
		\bar{ F}_1T(t)+\bar F_2\dot T(t)+\ddot T(t)=0,
		\end{equation}
		which verifies \eqref{eq:PDE-seperable}.
		To investigate the stability of  \eqref{eq:second-order}, consider $y^\top=[y^\top_1,y^\top_2]\triangleq[T^\top(t), \dot T^\top(t)]$. Thus, it follows that $\dot y=B y$, where
		\begin{equation}
		\label{eq:lyapanovEq}
		B= \begin{bmatrix}
		0&{I}\\[2.5pt]
		{ - {\bar F_1}}&{{-\bar F _2}}
		\end{bmatrix}
		\end{equation}
		Then, we define the Lyapunov function $V(y) = \frac{1}{2}y_1^\top{K_0}{y_1} + \frac{1}{2}y_2^\top M_0 {y_2}$. Hence, 
		\begin{equation}
		\begin{split}
		\dot V(y) &= y_2^\top{K_0}{y_1} - y_2^\top{M_0\bar F _1}{y_1} \\&- y_2^\top M_0{\bar F _2}{y_2}=y_2^\top({\bar{\mathcal{J}_0}-\bar{\mathcal{R}}_0}){y_2}\leq 0.
		\end{split}
		\end{equation}
		The largest invariant set inside $\mathcal{S}=\{y|\dot V(y)=0\}$ is $(y^\star_1,y^\star_2)=(0,0)$. Thus, it follows from LaSalle's Invariance principle that the origin $(y^\star_1,y^\star_2)$ is asymptotically stable.
		
		Then, due to the stability of \eqref{eq:second-order}, there exists a bound of the form
		\begin{equation}
		\label{eq:normT}
		\Vert T(t)\Vert \leq \Vert y(t)\Vert \leq \gamma, \quad \gamma>0.
		\end{equation}
		Since the system \eqref{eq:second-order} does not depend on $N$ (or  $\delta^k$ which represents the discretization step-size denoted in Definition \ref{definition3}), \eqref{eq:normT} holds for any $N\in\mathbb{N}$. Therefore \eqref{eq:seperation-Var} and \eqref{eq:normT} yield $\Vert \tilde q(\vec{z},t)\Vert\leq \gamma\prod\limits_{k=1}^{D}  \vert{X_k({z}^k)}\vert.$ 
		Accordingly,
		\begin{equation}
		\label{eq:fbound1}
		\Vert \tilde q(\vec{z},t)\Vert\leq \frac{\gamma}{\Vert T(t_0)\Vert}\Vert q(\vec{z},t_0)\Vert.
		\end{equation}
		Therefore, we conclude that
		\begin{equation}
		\label{eq:fbound3}
		\Vert \tilde q_i(t)\Vert\leq \frac{\gamma}{\Vert T(t_0)\Vert}\Vert{\tilde q_i(t_0)}\Vert \leq \frac{\gamma}{\Vert T(t_0)\Vert}\alpha.
		\end{equation}
		Recall that the bound \eqref{eq:fbound3} applies to any initial condition satisfying $\Vert\tilde q_i(t_0)\Vert<\alpha$ and holds for all ${i \in  \mathcal N}$ and any $N \in  \mathbb{N}$. Therefore, the bound \eqref{eq:fbound3} is also satisfied for $\max\limits_{i \in  \mathcal N} \Vert\tilde q_i(0)\Vert$. Hence, \eqref{eq:fbound3} is a uniform bound (independent of the number of agents). Moreover, the asymptotic stability of the system \eqref{eq:second-order}, together with \eqref{eq:sep-Var-norm}, implies $\Vert \tilde q(\vec{z},t)\Vert \to 0$. Therefore,  by \eqref{finitediffrence}, we conclude that the trajectories $\Vert\tilde q_i(t)\Vert$ approach zero as $t \to \infty$ for all ${i \in  \mathcal N}$ and ${\mathcal N \in  \mathbb{N}}$, which implies that the control objectives are achieved.	$\QEDA$ 
	\end{pf} 
\end{thm}
\section{Example}
\label{sec:p2}
In this section, we illustrate the applicability and performance of the proposed distributed controller for a network of spacecraft during the formation flying in a circular (or near-circular) reference orbit. To this end, each spacecraft is represented in the pH framework with the model derived in \cite[Chapter~3.1.2]{javanmardi2020spacecraft}. Furthermore, the spacecraft are modeled as point masses such that $M_0=I$. Hence, the position of each agent $i$ is denoted by $q_i(t)=[x_i,y_i,z_i]^\top \in {\mathbb{R}^{3}}$ in the local coordinates, known as the Hill frame. For further details on the model, we refer the reader to \cite{javanmardi2020spacecraft}. Moreover, we set the leader and follower spacecraft in the Medium Earth Orbit as a reference orbit with an altitude of $20\times 10^{3}$ km. 

For illustration purposes, we consider a network of $N=6$ spacecraft, where each agent (spacecraft) is in closed-loop with a controller of the form \eqref{controller-2}, and the communication graph is a 2-dimensional
mesh-like architecture like the one
described in Definition \ref{definition3}, with $r^1=3$ and $r^2=2$ depicted in Fig. \ref{Communication graph}(b). The control objective is as follows: each spacecraft $i$ must track a circular formation trajectory (known as the GCO formation in the SFF literature) in the space while maintaining the inter-agents differences $\Delta^{h}_{i,j}=[10,0,0]^\top$ and $\Delta^{v}_{i,j}=[0,20,0]^\top$ from its neighbors in the horizontal and vertical directions of the graph. In this connection, the reference trajectory $q^\ell_0(t)=[x(t),y(t),z(t)]^\top$ of the leader is considered as $x(t)=\frac{d^{ref}}{2}\cos (n_0t)$, $y(t)=-d^{ref}\sin(n_0t)$, $z(t)=\frac{\sqrt{3}}{2}d^{ref}\cos(n_0t)$, where the natural frequency of the reference orbit is $n_0=0.5307$, and the radius of the circular formation is $d^{ref}=5$ km. The control parameters used for simulations are $\bar{\mathcal{R}}_0=\mbox{diag}(34.4,42.3,10.59)$, $K_0=\mbox{diag}(30,30,20)$, and
\begin{equation*}
\bar{\mathcal{J}}_{0}=\begin{bmatrix}
0&1.0615&0\\-1.0615&0&0\\ 0&0&0\end{bmatrix}.
\end{equation*}
\begin{figure}
	\centering
	\includegraphics[width=0.35\textwidth]{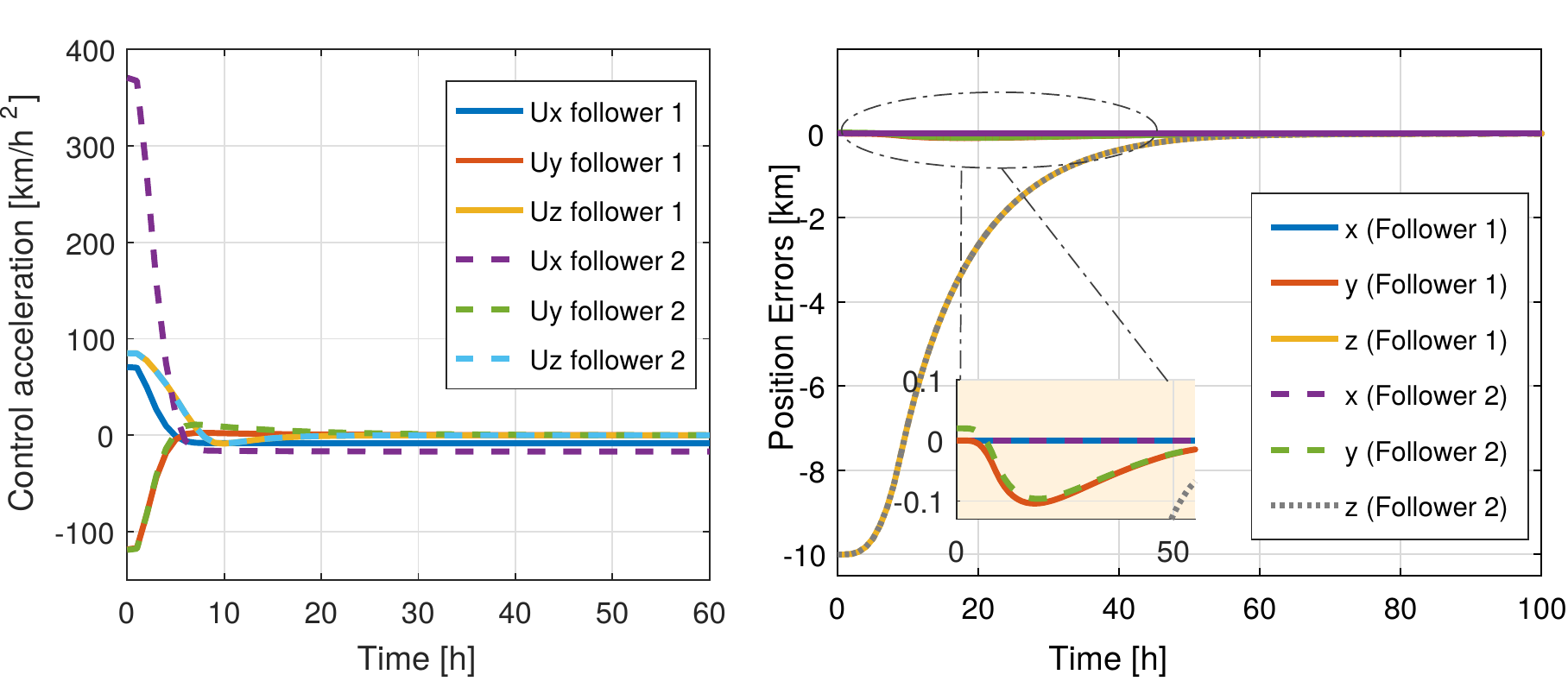}
	\centering
	\caption{Control acceleration and time responses of position errors of the follower 1 and 2.}
	\label{final4}
\end{figure}
\begin{figure}
	\centering
	\includegraphics[width=0.3\textwidth]{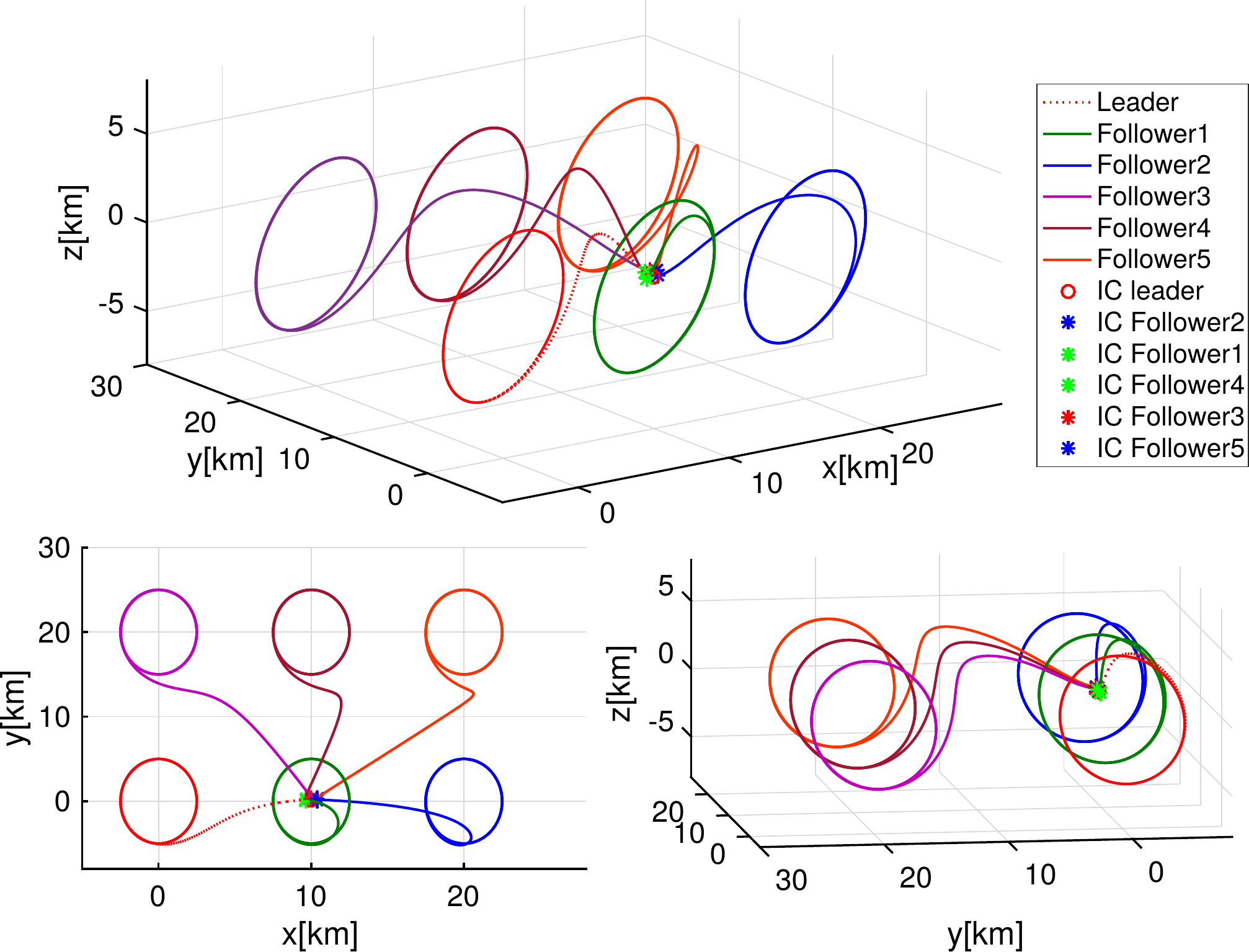}
	\centering
	\caption{Mission trajectories of the 6 agents in the network.}
	\label{final3}
\end{figure}
The position and velocity errors of the spacecraft converge to zero. In particular, Fig. \ref{final4} shows the errors of the follower 1 and 2. It is shown that amplification does not occur and we have a scalable performance . The three-dimensional trajectories of 6 agents are shown, from diverse viewing angles, in Fig. \ref{final3}. The mentioned plots corroborate that the control objective is achieved.

\section{Conclusion}
\label{sec:C7}
This paper proposed a systematic approach for deriving a distributed nonlinear controller based on the tIDA-PBC method for a large number of agents with mechanical structure.
We used the PDE-based method to analyze the scheme in terms of the stability and performance of the networked system.
It was shown that the proposed control law is a scalable scheme for the unidirectional architecture of the communication graph.
The proposed controller showed positive results in the simulation of a network of spacecraft.
\begin{ack}
The authors would like to thank Dr. Paul Kotyczka for valuable discussions during the preparation of the work.
\end{ack}

\bibliography{ifacconf}             

\begin{thebibliography}{16}
\providecommand{\natexlab}[1]{#1}
\providecommand{\url}[1]{\texttt{#1}}
\providecommand{\urlprefix}{URL }
\expandafter\ifx\csname urlstyle\endcsname\relax
  \providecommand{\doi}[1]{doi:\discretionary{}{}{}#1}\else
  \providecommand{\doi}{doi:\discretionary{}{}{}\begingroup
  \urlstyle{rm}\Url}\fi

\bibitem[{Barooah et~al.(2009)Barooah, Mehta, and
  Hespanha}]{barooah2009mistuning}
Barooah, P., Mehta, P.G., and Hespanha, J.P. (2009).
\newblock Mistuning-based control design to improve closed-loop stability
  margin of vehicular platoons.
\newblock \emph{IEEE Transactions on Automatic Control}, 54(9), 2100--2113.

\bibitem[{Besselink and Knorn(2018)}]{besselink2018scalable}
Besselink, B. and Knorn, S. (2018).
\newblock Scalable input-to-state stability for performance analysis of
  large-scale networks.
\newblock \emph{IEEE Control Systems Letters}, 2(3), 507--512.

\bibitem[{Dashkovskiy and Pavlichkov(2020)}]{dashkovskiy2020stability}
Dashkovskiy, S. and Pavlichkov, S. (2020).
\newblock Stability conditions for infinite networks of nonlinear systems and
  their application for stabilization.
\newblock \emph{Automatica}, 112, 108643.

\bibitem[{Duindam et~al.(2009)Duindam, Macchelli, Stramigioli, and
  Bruyninckx}]{duindam}
Duindam, V., Macchelli, A., Stramigioli, S., and Bruyninckx, H. (2009).
\newblock \emph{Modeling and control of complex physical systems: the
  port-Hamiltonian approach}.
\newblock Springer Science \& Business Media.

\bibitem[{Dunbar and Caveney(2011)}]{dunbar2011distributed}
Dunbar, W.B. and Caveney, D.S. (2011).
\newblock Distributed receding horizon control of vehicle platoons: Stability
  and string stability.
\newblock \emph{IEEE Transactions on Automatic Control}, 57(3), 620--633.

\bibitem[{Ghasemi et~al.(2013)Ghasemi, Kazemi, and Azadi}]{ghasemi2013stable}
Ghasemi, A., Kazemi, R., and Azadi, S. (2013).
\newblock Stable decentralized control of a platoon of vehicles with
  heterogeneous information feedback.
\newblock \emph{IEEE Transactions on Vehicular Technology}, 62(9), 4299--4308.

\bibitem[{Hao et~al.(2011)Hao, Barooah, and Mehta}]{hao2011stability}
Hao, H., Barooah, P., and Mehta, P.G. (2011).
\newblock Stability margin scaling laws for distributed formation control as a
  function of network structure.
\newblock \emph{IEEE Transactions on Automatic Control}, (4), 923--929.

\bibitem[{Javanmardi et~al.(2020)Javanmardi, Yaghmaei, and
  Yazdanpanah}]{javanmardi2020spacecraft}
Javanmardi, N., Yaghmaei, A., and Yazdanpanah, M.J. (2020).
\newblock Spacecraft formation flying in the port-hamiltonian framework.
\newblock \emph{Nonlinear Dynamics}, 1--19.

\bibitem[{Knorn et~al.(2015)Knorn, Donaire, Ag{\"u}ero, and
  Middleton}]{knorn2015scalability}
Knorn, S., Donaire, A., Ag{\"u}ero, J.C., and Middleton, R.H. (2015).
\newblock Scalability of bidirectional vehicle strings with static and dynamic
  measurement errors.
\newblock \emph{Automatica}, 62, 208--212.

\bibitem[{Meurer(2012)}]{meurer2012control}
Meurer, T. (2012).
\newblock \emph{Control of higher--dimensional PDEs: Flatness and backstepping
  designs}.
\newblock Springer Science \& Business Media.

\bibitem[{Ortega et~al.(2002)Ortega, Spong, G{\'o}mez-Estern, and
  Blankenstein}]{ortega2002stabilization}
Ortega, R., Spong, M.W., G{\'o}mez-Estern, F., and Blankenstein, G. (2002).
\newblock Stabilization of a class of underactuated mechanical systems via
  interconnection and damping assignment.
\newblock \emph{IEEE transactions on automatic control}, 47(8), 1218--1233.

\bibitem[{Ren and Sorensen(2008)}]{ren2008distributed}
Ren, W. and Sorensen, N. (2008).
\newblock Distributed coordination architecture for multi-robot formation
  control.
\newblock \emph{Robotics and Autonomous Systems}, 56(4), 324--333.

\bibitem[{Tsolakis and Keviczky(2021)}]{tsolakis2021distributed}
Tsolakis, A. and Keviczky, T. (2021).
\newblock Distributed {IDA}-{PBC} for a class of nonholonomic mechanical
  systems.
\newblock \emph{IFAC-PapersOnLine}, 54(14), 275--280.

\bibitem[{Valk and Keviczky(2018)}]{valk2018distributed}
Valk, L. and Keviczky, T. (2018).
\newblock Distributed control of heterogeneous underactuated mechanical
  systems.
\newblock \emph{IFAC-PapersOnLine}, 51(23), 325--330.

\bibitem[{Yaghmaei and Yazdanpanah(2017)}]{yaghmaei2017trajectory}
Yaghmaei, A. and Yazdanpanah, M.J. (2017).
\newblock Trajectory tracking for a class of contractive port hamiltonian
  systems.
\newblock \emph{Automatica}, 83, 331--336.

\bibitem[{Yazdanian and Mehrizi-Sani(2014)}]{yazdanian2014distributed}
Yazdanian, M. and Mehrizi-Sani, A. (2014).
\newblock Distributed control techniques in microgrids.
\newblock \emph{IEEE Transactions on Smart Grid}, 5(6), 2901--2909.

\end{thebibliography}
                                                   







\end{document}